\documentclass[a4paper,11pt]{article}
\pdfoutput=1
\usepackage{jheppub}
\usepackage{amsmath,amssymb,amsthm,mathrsfs}
\usepackage{booktabs}
\let\jhepbibliography\thebibliography
\renewcommand{\thebibliography}[1]{\jhepbibliography{#1}\setlength{\itemsep}{3pt}}

\newcommand{\Tr}{\operatorname{Tr}}
\newcommand{\ES}{\mathrm{ES}}
\newcommand{\Lm}{\mathcal{L}^-}
\newcommand{\Bm}{\mathcal{B}^-}
\newcommand{\ad}{\operatorname{ad}}
\makeatletter\@ifundefined{notoc}{}{\notoc}\makeatother

\title{Remark on residual solutions in the $KBc$ algebra}

\author[a,b]{Ilmar Gahramanov}

\affiliation[a]{Department of Physics, Bo\u{g}azi\c{c}i University,\\
34342 Bebek, Istanbul, T\"urkiye}
\affiliation[b]{Center for Mathematics and its Applications, Khazar University,\\
Mehseti St.~41, AZ1096, Baku, Azerbaijan}

\emailAdd{ilmar.gahramanov@bogazici.edu.tr}

\abstract{In this paper we study residual solutions of open bosonic string field theory in the $KBc$ algebra with the auxiliary identities $(\partial^mc)^2=0$ of the ghost field. These solutions are not covered by the classification based on the defining relations alone. We show that the operator which controls the cohomology on the residual family depends only on $s=\gamma_1+\gamma_2$. At positive level $n$, cohomology can appear only on the lines $s=\pm n$. We compute it through level four and find one class at every point of each line $s=-3$ and $s=-4$. These classes give exact polynomial deformations of the residual family. At level six on $s=-6$, the cohomology instead vanishes generically and has dimension two at $(-1,-5)$ and $(-5,-1)$. We also compute the failure of the standard generator redefinition to preserve the auxiliary identities, and discuss the regularization of an explicit residual solution.}

\keywords{String Field Theory, Tachyon Condensation, Bosonic Strings}

\begin{document}
\hypersetup{pageanchor=false}
\maketitle
\flushbottom
\newpage
\setcounter{page}{1}
\hypersetup{pageanchor=true}

\section{Introduction}

In 2005 Schnabl found the first analytic solution for tachyon condensation in Witten's open string field theory \cite{Witten,Schnabl}. In 2009 Erler and Schnabl found a simpler solution \cite{ErlerSchnabl}; a short account is given in \cite{ErlerTMP}. Since then the $KBc$ algebra has become a standard tool for constructing analytic solutions. Reviews can be found in \cite{BonoraIntroduction,BonoraProspects,ErlerLectures}, and a broader introduction to string field theory is given in \cite{ArefevaTMP}. Many solutions in this algebra contain the identity string field. Their interpretation requires care \cite{KishimotoTakahashi,Zeze}, and identity-based solutions continue to appear in studies of open string renormalization group flow \cite{Hristov1,Hristov2}.

In 2012 Erler classified solutions which admit an $\Lm$ level expansion and satisfy the equations of motion by the defining relations of the $KBc$ algebra alone \cite{ErlerId}. Under these assumptions there are six gauge orbits: the perturbative vacuum, the tachyon vacuum and four residual solutions. The residual solutions are singular from the point of view of the identity string field.

The ghost field satisfies additional relations, $(\partial^mc)^2=0$ for $m\ge1$, which do not follow from the defining relations. These are the auxiliary identities of \cite{ErlerId}. With them, the four residual solutions become points of a two-parameter family. The classification does not cover solutions which need these identities. It was also observed in \cite{ErlerId} that the cohomology of the residual family may be enhanced at special points, with a possible connection to multibrane solutions \cite{MurataSchnabl}.

Here we study the cohomology of this family. We first give a criterion for imposing the auxiliary identities: only the part antisymmetric in the ghost derivatives survives. We then show that the operator which controls the cohomology depends only on $s=\gamma_1+\gamma_2$. At positive level $n$, a class can occur only when $s=\pm n$. Through level four the positive lines have no cohomology. On the negative lines we recover the known level-two class at the ghost brane point and find one class at every point of each line $s=-3$ and $s=-4$. These classes give exact polynomial solutions without higher-order corrections.

We also examine level six on $s=-6$. Here the cohomology vanishes at generic points and has dimension two at $(-1,-5)$ and $(-5,-1)$. Thus the classes found on the whole lines at levels three and four do not imply the same behavior at higher levels. Finally, we compute explicitly how the generator redefinition used in \cite{ErlerId} fails to preserve the auxiliary identities.

Our interest in this question started from the family $\psi_b=c(1+K)Bc\big((1+K)^{-1}+b\big)$, discussed in the lecture \cite{Gahramanov}. This is a special case of the two-parameter solution in section~2.3 of \cite{Arroyo}. It also appears in equation~(86) of \cite{BonoraTolla}, with $\lambda=1$ and $\alpha=1+b$. We return to this family and its regularization in section~7.

The cohomology calculations and the new polynomial solutions are statements in the algebra with auxiliary identities. Their interpretation as regular string backgrounds, and the definition of their energy, require further analysis.\footnote{The algebraic calculations were checked with a computer implementation of the $KBc$ algebra. The accompanying code includes an independent check in an ordered monomial basis.}

The rest of the paper is organized as follows. In section~2 we review the $KBc$ algebra and the auxiliary identities. Section~3 describes the residual family, and section~4 studies its cohomology through level four. In section~5 we use the cohomology classes to construct new solutions. Section~6 discusses the generator redefinition, and section~7 examines the explicit family $\psi_b$ and its regularization. We conclude in section~8. Appendix~\ref{app:ellwood} gives the closed string overlap, and appendix~\ref{app:levelsix} contains the level-six calculation.

\section{The \texorpdfstring{$KBc$}{KBc} algebra and auxiliary identities}
\label{sec:algebra}

\subsection{The \texorpdfstring{$KBc$}{KBc} algebra}
The string fields $K$, $B$ and $c$ have Grassmann parity even, odd and odd, respectively. Their defining relations are
\begin{equation}
\{B,c\}=1,\qquad B^2=c^2=0,\qquad [K,B]=0,\qquad QB=K,\qquad QK=0,\qquad Qc=cKc .
\label{eq:kbc}
\end{equation}
We suppress the star product symbol and write $\Omega^t=e^{-tK}$. The BRST operator around $\psi$ is $Q_\psi X=QX+\psi X-(-1)^{|X|}X\psi$. For any function $X(K)$, the defining relations give
\begin{equation}
cBc=c,\qquad BcXBc=BXc,\qquad Q(Bc)=cKBc,\qquad (Bc)^2=Bc .
\label{eq:ids}
\end{equation}
Thus $P=Bc$ and $\bar P=cB=1-P$ are complementary projectors. For a function
$G(K)$, set $M=K/(1-G)$. Then
\begin{equation}
\psi_G=cMBcG
\label{eq:okawa}
\end{equation}
is the left form of Okawa's solution \cite{Okawa}. The case $G=(1+K)^{-1}$ gives the Erler--Schnabl
solution $\psi_\ES$.\footnote{The interpretation of (\ref{eq:okawa}) as a tachyon vacuum or as a pure
gauge solution requires regularity of the functions involved, not only conditions at $K=0$.} We
normalize the energy so that $E[\psi_\ES]=-1/(2\pi^2)$.

\subsection{Auxiliary identities}
\label{sec:aux}
Write $\partial X=[K,X]$ and $c_j=\partial^jc$. In addition to the defining relations, the ghost field obeys
\begin{equation}
c_j^2=0\quad(j\ge1),\qquad\text{and hence}\qquad \{c_i,c_j\}=0\quad(i,j\ge0) .
\label{eq:aux}
\end{equation}
The anticommutators follow by differentiating the squared identities, including $c^2=0$. More explicitly, induction on $i+j$ relates adjacent pairs of indices, while the middle pair is fixed by a squared identity or its derivative. These relations are independent of (\ref{eq:kbc}) \cite{ErlerId}. They allow us to write every polynomial string field as
\begin{equation}
\sum a_{r;i_1\ldots i_p;\eta}\,K^r c_{i_1}\cdots c_{i_p} B^\eta,
\qquad r\ge0,\quad i_1<\cdots<i_p,\quad \eta\in\{0,1\},
\label{eq:pbw}
\end{equation}
using the rules
\begin{equation}
[K,c_j]=c_{j+1},\qquad \{B,c_j\}=\delta_{j0},\qquad
Qc_j=\sum_{r=0}^j\binom jr c_r c_{j-r+1}.
\label{eq:rewrite}
\end{equation}
A monomial in (\ref{eq:pbw}) has ghost number $p-\eta$ and level
\begin{equation}
\ell=r+\sum_{a=1}^p(i_a-1)+\eta ,
\label{eq:level}
\end{equation}
At fixed ghost number and level the state space is therefore finite-dimensional. We compute cohomology in this polynomial algebra.\footnote{Inverse powers of $K$ are excluded.
Otherwise $Q(B/K)=1$ would make the cohomology trivial even at the perturbative vacuum.}

\subsection{A simple criterion}
The auxiliary identities can also be imposed in a polynomial representation. Associate the word
\begin{equation}
f_0(K)\,c\,f_1(K)\,c\cdots c\,f_m(K)\,[B],
\end{equation}
with a polynomial $F(x_0,\ldots,x_m)$, where $x_i$ is the argument of $f_i$, and define
$g_i=x_i-x_{i-1}$. Differentiating the $i$-th ghost multiplies $F$ by $-g_i$. The relations
$\{c_p,c_q\}=0$ therefore remove the part symmetric under the exchange of two neighbouring gaps.
Since neighbouring transpositions generate $S_m$, we obtain
\begin{equation}
F=0\ \text{in the auxiliary quotient}\quad\Longleftrightarrow\quad
\sum_{\pi\in S_m}\operatorname{sgn}(\pi)\,F(x_0;g_{\pi(1)},\ldots,g_{\pi(m)})=0 .
\label{eq:antisym}
\end{equation}
Thus only the part antisymmetric in the ghost derivatives survives. This criterion reproduces the cohomology around the four residual solutions
given in table~2 of \cite{ErlerId} at levels $-1$ to $2$.

\section{The residual family}
\label{sec:moduli}

With the auxiliary identities, the level-zero residual solutions belong to the family \cite{ErlerId}
\begin{equation}
\Psi_0=\gamma_1\,cK+\gamma_2\,Kc-(1+\gamma_1+\gamma_2)\,cKBc,\qquad s=\gamma_1+\gamma_2 .
\label{eq:psi0}
\end{equation}
The points $(0,0)$, $(-1,0)$, $(0,-1)$ and $(-1,-1)$ are the residual perturbative vacuum, the
residual tachyon vacuum, the residual conjugate tachyon vacuum and the ghost brane solution
\cite{MNT}. Only these four points satisfy the equations of motion by (\ref{eq:kbc}) alone. The remaining points require the auxiliary identities. We write $D=Q_{\Psi_0}$.

Following \cite{ErlerId}, introduce the odd derivation $\beta=\frac12\Bm$, with $\beta K=B$ and
$\beta B=\beta c=0$. It is well defined in the auxiliary quotient, since $\beta c_1=1$ and $\beta c_j=0$ for $j\ne1$, and hence it preserves the anticommutation relations. Using $\beta(cK)=-cB$, $\beta(Kc)=Bc$ and $\beta(cKBc)=0$, we obtain
$\beta\Psi_0=\gamma_2-s\,cB$. Consequently,
\begin{equation}
\mathscr L\equiv\{D,\beta\}=\frac12\Lm-s\,\ad_{cB} .
\label{eq:Lpsi}
\end{equation}
The operator therefore depends only on $s$. If $D\varphi=0$ and
$\mathscr L\varphi=h\varphi$ with $h\ne0$, then $\varphi=D(\beta\varphi/h)$, so the cohomology lies in the
kernel of $\mathscr L$. Since $cB$ is a projector, $\ad_{cB}$ has the eigenvalues $e=-1,0,+1$ on
$PX\bar P$, $PXP+\bar PX\bar P$ and $\bar PXP$. Moreover $D(cB)=0$, so $D$ preserves these sectors
and the level. A cohomology class at level $m$ must therefore satisfy
\begin{equation}
m=s\,e .
\label{eq:kernel}
\end{equation}
At ghost number one the possible classes are the marginal directions at $m=0$; the
state $c$ at $m=-1$, $e=+1$, which appears only for $s=-1$; and states at positive level $m=n$, with
$e=-1$ on the lines $s=-n$ or with $e=+1$ on the lines $s=n$.

At level zero, the tangent vectors $v_1=cK-cKBc$ and $v_2=Kc-cKBc$ are closed and
independent. The only other state with ghost number one is $cKBc$, and using (\ref{eq:rewrite}) we
find
\begin{equation}
D(cKBc)=s\,Kc_0c_1-\gamma_1c_0c_2 .
\label{eq:levelzero}
\end{equation}
Both states with ghost number zero, $1$ and $Bc$, are annihilated by $D$. Hence the level-zero
cohomology is two-dimensional everywhere except at $(0,0)$, where the residual perturbative vacuum
has the additional class $cKBc$ \cite{ErlerId}. At level $-1$ the state $c$ is closed only for
$s=-1$, and then it is not exact.

\section{Cohomology at positive level}
\label{sec:coh}

For $1\le n\le4$, the sector $e=+1$ on $s=n$ has no closed states of ghost number one. We therefore consider
$e=-1$ on $s=-n$, and set $\gamma_1=a$, $\gamma_2=-n-a$. In this sector the states with
ghost number one at level $n$ are $K^rc_ic_jB$ with $1\le i<j$ and $r+i+j=n+1$, and the states with
ghost number zero are $K^{n-j}c_jB$ with $1\le j\le n$. Since $D$ preserves the level and the
sectors, exactness in this sector is the same as exactness in the whole algebra.

This sector has no ghost number one states at level one. At level two its only state is $u=c_1c_2B$. We find
\begin{equation}
Du=0,\qquad D(Kc_1B)=-(a+1)\,u,\qquad D(c_2B)=0 .
\label{eq:n2maps}
\end{equation}
A class therefore survives only at $a=-1$, the ghost brane point \cite{MNT,ErlerId}.

At level three a basis is $u=Kc_1c_2B$, $v=c_1c_3B$. The differential is
\begin{equation}
Du=Dv=0,\qquad D(K^2c_1B)=-(2a+3)\,u+(a+2)\,v,\qquad D(Kc_2B)=D(c_3B)=0 .
\label{eq:n3maps}
\end{equation}
Since $2a+3$ and $a+2$ never vanish together, the image is one-dimensional for every $a$. The state
\begin{equation}
\varphi_3=2u-v=(2Kc_1c_2-c_1c_3)B
\label{eq:phi3}
\end{equation}
is closed and is not exact for any $a$, since
$\det\begin{pmatrix}-(2a+3)&a+2\\2&-1\end{pmatrix}=-1$.

At level four, write
\begin{equation}
u_1=K^2c_1c_2B,\quad u_2=Kc_1c_3B,\quad u_3=c_1c_4B,\quad u_4=c_2c_3B,\quad w=c_1c_2c_3B .
\end{equation}
On the ghost number one states,
\begin{equation}
Du_1=Du_2=-(a+3)\,w,\qquad Du_3=-2w,\qquad Du_4=0 ,
\label{eq:n4closed}
\end{equation}
while on the ghost number zero states,
\begin{equation}
\begin{aligned}
D(K^3c_1B)&=-3(a+2)\,u_1+(3a+8)\,u_2-(a+3)\,u_3,\\
D(K^2c_2B)&=D(Kc_3B)=(a+3)\,u_4,\qquad D(c_4B)=2u_4 .
\end{aligned}
\label{eq:n4image}
\end{equation}
The kernel is therefore three-dimensional and the image two-dimensional for every $a$. The state
\begin{equation}
\varphi_4=u_1-u_2=(K^2c_1c_2-Kc_1c_3)B
\label{eq:phi4}
\end{equation}
is closed. For $a\ne-3$ it is not exact, since it has no $u_3$ component, and at $a=-3$ the first
image vector becomes $3u_1-u_2$, which is independent of $u_1-u_2$. Thus $\varphi_4$ is not exact anywhere on the
line. We collect these results in table~\ref{tab:coh}.

\begin{table}[t]
\centering
\begin{tabular}{@{}cccc@{}}
\toprule
$s=-n$, level $n$, $e=-1$ & $\dim\ker D$ & $\dim\operatorname{im}D$ & $\dim H^1$\\
\midrule
$n=1$ & 0&0&0\\
$n=2$, $a\ne-1$ &1&1&0\\
$n=2$, $a=-1$ &1&0&1\\
$n=3$, all $a$ &2&1&1\\
$n=4$, all $a$ &3&2&1\\
\bottomrule
\end{tabular}
\caption{Ghost number one cohomology in the sector $e=-1$ at level $n$ on the line
$\gamma_1+\gamma_2=-n$, with $\gamma_1=a$.}
\label{tab:coh}
\end{table}

At level six on $s=-6$, the cohomology is instead zero at generic points and two-dimensional at $(-1,-5)$ and $(-5,-1)$. The calculation is given in appendix~\ref{app:levelsix}.

\section{New solutions}
\label{sec:new}

The classes at levels two, three and four give exact solutions. At a point $\Psi_0$ on
$s=-n$, take
\begin{equation}
\Psi=\Psi_0+\epsilon\,\varphi_n ,
\label{eq:new}
\end{equation}
where $\epsilon$ is a number and $\varphi_n$ is one of the states found in the previous section, with
$\varphi_2=c_1c_2B$ at the ghost brane point. Since $\Psi_0$ is a solution, the equation of motion gives
\begin{equation}
Q\Psi+\Psi^2=\epsilon\,D\varphi_n+\epsilon^2\,\varphi_n^2 .
\end{equation}
The first term is zero because $D\varphi_n=0$. Also $\varphi_n^2=0$: all ghosts in $\varphi_n$ carry derivatives, so the final $B$ in the first factor can be moved through the second factor without a contraction, giving $B^2=0$. Equivalently,
$\varphi_n$ lies in $PX\bar P$, and $\bar PP=0$. Therefore (\ref{eq:new}) is a solution for every value of
$\epsilon$.

On the line $s=-3$ the solution is
\begin{equation}
\Psi=a\,cK-(3+a)\,Kc+2\,cKBc+\epsilon\,\big(2K\partial c\,\partial^2c-\partial c\,\partial^3c\big)B ,
\end{equation}
and on the line $s=-4$
\begin{equation}
\Psi=a\,cK-(4+a)\,Kc+3\,cKBc+\epsilon\,\big(K^2\partial c\,\partial^2c-K\partial c\,\partial^3c\big)B ,
\end{equation}
where $a$ can be any number. At the ghost brane point, the known level-two class \cite{ErlerId} similarly gives
\begin{equation}
\Psi=-cK-Kc+cKBc+\epsilon\,\partial c\,\partial^2c\,B .
\end{equation}
Since $\varphi_n$ is not exact, these are nontrivial deformations of $\Psi_0$.\footnote{This is a statement
about infinitesimal deformations at a fixed point of the family. In fact, the formal gauge transformation $U=1+\lambda Bc$, $\lambda\ne-1$, leaves $\Psi_0$ fixed and sends $\epsilon\mapsto\epsilon/(1+\lambda)$. Thus nonzero values of $\epsilon$ are formally gauge related; their physical interpretation would require a specified class of admissible string fields and gauge transformations.} These deformations require the auxiliary identities; for a nonzero deformation the defining relations (\ref{eq:kbc}) alone do not give the equations of motion.

More general formal power-series deformations along these classes are also unobstructed. At order $k\ge2$ the obstruction lies at level $kn$, where $\mathscr L$ has the eigenvalue
$n(k+e)>0$. Therefore a closed obstruction $S_k$ is exact, and it can be removed by
\begin{equation}
\chi_k=-\sum_e\frac{\beta(P_eS_k)}{n(k+e)},
\end{equation}
Here $P_e$ projects onto the sector with eigenvalue $e$. For the representatives above, $\varphi_n^2=0$ and all higher-order corrections vanish.

These solutions are polynomial in $K$ and are therefore identity-based. The positive-level terms increase their singularity. Computing an action or a boundary state requires a suitable regularization.

\section{The generator redefinition}
\label{sec:puzzle}

The energy prescription in \cite{ErlerId} uses the generator redefinition
\begin{equation}
\widehat K=\frac K{1+\delta K},\qquad \widehat B=\frac B{1+\delta K},\qquad \widehat c=c(1+\delta K)Bc ,
\label{eq:hat}
\end{equation}
This preserves (\ref{eq:kbc}) and commutes with $Q$. The energy argument also assumes that a BRST-exact identity-like term has zero overlap. The redefined generators need not obey the auxiliary identities \cite{ErlerId}. Expanding to first order,
$\widehat K=K-\delta K^2+O(\delta^2)$ and $\widehat c=c+\delta\,cKBc$, and using (\ref{eq:rewrite}) we find
\begin{equation}
(\widehat\partial\widehat c)^2=-\delta\,c_1c_2+O(\delta^2)\ne0 ,
\label{eq:hatdefect}
\end{equation}
where $\widehat\partial=[\widehat K,\cdot\,]$. The redefinition (\ref{eq:hat}) therefore preserves the
$KBc$ algebra but does not descend to the auxiliary quotient. For a solution which needs these
identities, the redefined field solves the equations of motion only if the redefinition maps the
defect $Q\Psi+\Psi^2$ back into the auxiliary ideal, and this has to be checked case by case.

This explains an obstruction behind the energy puzzle in \cite{ErlerId}. The energy rule found there is
proportional to $s=\gamma_1+\gamma_2$, while the auxiliary identities connect the residual solutions by
marginal directions, which suggests that their energies should be equal. Since the regularization
behind the energy rule does not respect the auxiliary identities, the redefinition is guaranteed to preserve the equations of motion only at
the four points which satisfy the defining relations. Even there, the energy assignment still depends on the assumed trace prescription. It cannot simply be extended along the
family. An energy calculation on this family requires a regularization
compatible with the auxiliary identities.

\section{An explicit example}
\label{sec:family}

For the family discussed in the introduction, set $G=(1+K)^{-1}$ and $M=1+K$. Then
\begin{equation}
\psi_b=c(1+K)Bc\left(\frac1{1+K}+b\right)=\psi_\ES+b\,\Phi,\qquad \Phi=c(1+K)Bc .
\label{eq:psib}
\end{equation}
Using (\ref{eq:ids}), we find $Q\Phi=cKc$, $\Phi^2=\Phi\psi_\ES=0$ and $\psi_\ES\Phi=-cKc$,
so $\psi_b$ solves the equations of motion for every $b$.\footnote{We correct here a formula
in \cite{Gahramanov}. For the ansatz $\psi=c(1+K)BcF$ with $F=(a+\beta K)/(1+K)$ the product should read
$\psi^2=(\beta-a)\,cKcF$; the condition $a-\beta=1$ stated there is correct.} Expanding around $K=0$,
$\psi_b=(1+b)c-cK+(1+b)cKBc+\cdots$, so for $b\neq-1$ it is in the tachyon vacuum orbit and for $b=-1$ it
is in the residual tachyon vacuum orbit: its level-zero term is $-cK$ \cite{ErlerId}. The gauge transformation
$U=1+\lambda Bc$, with $\lambda\ne-1$, acts as $1+b\mapsto(1+\lambda)(1+b)$, so $b=-1$ is a fixed point.

At $b=-1$ there is no homotopy field in this $KBc$ algebra. A field of ghost number minus one has the form $BF(K)$, and
\begin{equation}
Q_{\psi_{-1}}(BF)=-\big[cKF+FMc(G-1)\big]B ,
\label{eq:residualhomotopy}
\end{equation}
and the right-hand side vanishes when multiplied by $B$ from the right, while $1\cdot B\ne0$. For
functions with suitable Laplace representations the closed string overlap \cite{Ellwood} is (see
appendix~\ref{app:ellwood})
\begin{equation}
W[f_0cf_1]=-(f_1f_0)'(0),\qquad W[f_0cf_1cf_2B]=-f_1'(0)\,(f_2f_0)(0),
\label{eq:ellwood}
\end{equation}
which gives $W[\psi_G]=-M(0)G'(0)$ for regular Okawa solutions. Formally, the energy and the overlap
of $\psi_b$ are those of the tachyon vacuum for every $b$ \cite{Arroyo,Gahramanov,ErlerId}, but for
$b\ne0$ the traces contain identity-like terms, and one needs a prescription to define them. The interpretation of singular string fields in terms of generalized states is discussed in \cite{BonoraGiaccari}. For related discussions of identity-based solutions and the distinction between regular and singular gauge transformations, see \cite{BonoraTolla}.

To regularize the gauge parameter, take $R_\mu=(1+\mu K)^{-1}$, $\mu>0$, and
$U_b^\mu=1+bBcR_\mu$. Then
\begin{equation}
\psi_b^{R_\mu}=(U_b^\mu)^{-1}(Q+\psi_\ES)\,U_b^\mu=f^{-1}\psi_{\widetilde G}\,f,\qquad
f=1+bR_\mu,\qquad \widetilde G=\frac{G+bR_\mu}{f} ,
\label{eq:conj}
\end{equation}
Thus the regularized field is an Okawa solution conjugated by a function of $K$. For $b>-1$ the pole of $f^{-1}$ lies on the
negative real axis, so the functions in the gauge parameter and its inverse admit Laplace representations supported at nonnegative wedge widths. With the usual wedge-state prescription this gives a regular gauge transformation of the tachyon vacuum. Its coefficientwise limit as $\mu\to0$ is $\psi_b$; this does not establish continuity of singular traces.\footnote{The conditions $R(0)=1$ and $R(\infty)=0$ alone are
not enough. For $R=e^{-\delta K}$ the poles of $(1+bR)^{-1}$ at $K=(\log b+i(2j+1)\pi)/\delta$ lie in the
right half-plane when $b>1$.} For $b=-1$, however, $f_\mu=\mu K/(1+\mu K)$ vanishes at $K=0$ and
$\widetilde G_\mu=-(1-\mu)G/\mu$, so (\ref{eq:conj}) is a singular conjugation. Its coefficientwise limit in the localized algebra can be computed
explicitly:
\begin{equation}
\lim_{\mu\to0}\psi_{-1}^{R_\mu}=-K^{-1}c\,(1+K)K\,Bc\,\frac K{1+K}\equiv\psi_* ,
\qquad (\psi_*-\psi_{-1})B=K^{-1}[K,c]\,KB\ne0 .
\label{eq:actual-limit}
\end{equation}
The regularized family therefore does not return to the residual field. For traces which are defined at finite cutoff, one can see where
the difficulty comes from: for $S_N=\sum_{j=0}^NR^j$ we have $\Tr[S_NX(1-R)]=\Tr X-\Tr[R^{N+1}X]$, and
the boundary term has to be controlled before an energy can be assigned to the singular conjugation.

For $b<-1$, a real continuous regulator with $R(0)=1$ and $R(\infty)=0$ makes $1+bR$ vanish at some
$K_*>0$, where the numerator $G(K_*)-1$ of $\widetilde G$ does not vanish. Thus this regulated real projector ansatz
cannot connect the two signs of $1+b$ by a regular transformation, although more general gauge
parameters are not excluded. For the role of $K=0$ and $K=\infty$ in singular solutions see
\cite{HataKojita,HataKojita2}.

\section{Conclusions}
\label{sec:concl}

In this paper we studied residual solutions of the $KBc$ algebra with the auxiliary identities of the ghost field. The operator which controls the cohomology depends only on $s=\gamma_1+\gamma_2$ and restricts possible classes at positive level $n$ to $s=\pm n$. Through level four we find no cohomology on the positive lines. On the negative lines there is one class at every point of $s=-3$ and $s=-4$, while at level two the class survives only at the ghost brane point. The explicit representatives give exact polynomial deformations of the residual solutions.

The calculation in appendix~\ref{app:levelsix} shows a different behavior at level six. On $s=-6$ the cohomology vanishes generically and has dimension two at $(-1,-5)$ and $(-5,-1)$. A general description must therefore account for both classes which persist along a whole line and enhancement at isolated points.

We also showed that the standard generator redefinition breaks the auxiliary identities already at first order. The energy prescription based on this redefinition cannot be extended directly to the residual family. For the explicit family $\psi_b$, regularizing the gauge parameter gives a regular gauge transformation for $b>-1$ with the stated wedge-state prescription. At $b=-1$ the coefficientwise limit differs from the original residual field.

The main open question is the cohomology at arbitrary level. It would also be useful to find a regularization compatible with the auxiliary identities and to determine whether any of these formal solutions admit a boundary-state or multibrane interpretation \cite{MurataSchnabl}. Other realizations of the $KBc$ algebra, such as those in \cite{HataTakedaYoshinaka}, should be checked against the same identities.

\acknowledgments

I am deeply grateful to Loriano Bonora, with whom this project was started and from whom I learned this subject. I am also grateful to T.~Erler for correspondence on the family (\ref{eq:psib}).
I acknowledge support from the Istanbul Integrability and Stringy Topics Initiative (IStringy).
Algebraic checks and the preparation of the draft were assisted by Claude (Anthropic), and ChatGPT (OpenAI).

\appendix
\section{The closed string overlap}
\label{app:ellwood}

Let $V$ be an on-shell closed string vertex operator of ghost number two and weight $(0,0)$, inserted at the midpoint. Normalize its correlator by $\langle Vc(x)\rangle_{C_L}=(L/\pi)\,\mathcal C_V$. If $F$ has a Laplace representation with an integrable first moment, then
\begin{equation}
F(K)=\int_0^\infty dt\,\widetilde F(t)\,e^{-tK},\qquad \Tr_V[cF]=-\frac{\mathcal C_V}{\pi}\,F'(0),
\end{equation}
Cyclicity gives the first formula in (\ref{eq:ellwood}). For the second, choose the orientation of the
$B$ line so that, for widths $t_i\ge0$,
\begin{equation}
\Tr_V[\Omega^{t_0}c\,\Omega^{t_1}c\,\Omega^{t_2}B]=\frac{\mathcal C_V}{\pi}\,t_1 .
\end{equation}
Integrating over the widths we obtain
\begin{equation}
\Tr_V[f_0cf_1cf_2B]=-\frac{\mathcal C_V}{\pi}\,f_1'(0)\,(f_2f_0)(0).
\end{equation}
This sign agrees with $Q(BcF)=cKF-cKcFB$ and the vanishing overlap of BRST exact fields. We
define $W$ as the coefficient of $\mathcal C_V/\pi$. For regular Okawa solutions, writing $Bc=1-cB$ gives
$W[\psi_G]=-(MG)'(0)+M'(0)G(0)=-M(0)G'(0)$. These formulas assume that the traces converge; they
cannot be used to cancel singular factors inside otherwise undefined traces.

\section{Cohomology at level six}
\label{app:levelsix}

Here we give the level-six calculation on $s=-6$. Set $\gamma_1=a$ and $\gamma_2=-6-a$. In the sector $e=-1$, bases at ghost numbers zero, one and two are
\begin{equation}
h_j=K^{6-j}c_jB,\qquad j=1,\ldots,6,
\end{equation}
\begin{equation}
\begin{aligned}
(u_1,u_2,u_3)&=(K^4c_1c_2B,\ K^3c_1c_3B,\ K^2c_1c_4B),\\
(u_4,u_5,u_6)&=(Kc_1c_5B,\ c_1c_6B,\ K^2c_2c_3B),\\
(u_7,u_8,u_9)&=(Kc_2c_4B,\ c_2c_5B,\ c_3c_4B),
\end{aligned}
\end{equation}
and
\begin{equation}
(w_1,w_2,w_3,w_4)=(K^2c_1c_2c_3B,\ Kc_1c_2c_4B,\ c_1c_2c_5B,\ c_1c_3c_4B),
\end{equation}
respectively. The matrices of $D$ in these ordered bases are
\begin{equation}
D_0=\begin{pmatrix}v&0_{5\times5}\\0_{4\times1}&-A\end{pmatrix},
\qquad D_1=\begin{pmatrix}A&0_{4\times4}\end{pmatrix},
\label{eq:n6blocks}
\end{equation}
where
\begin{equation}
v=\begin{pmatrix}
-5a-15\\10a+40\\-10a-45\\5a+24\\-a-5
\end{pmatrix},\qquad
A=\begin{pmatrix}
-6a-26&-3a-12&-2&0&0\\
4a+19&0&-2a-9&-5&0\\
-a-5&0&0&-a-5&-9\\
0&a+5&a+5&0&-5
\end{pmatrix}.
\label{eq:n6matrix}
\end{equation}
The relation $Av=0$ gives $D_1D_0=0$, and $v$ is nonzero for every $a$. The greatest common divisor of the nonzero $4\times4$ minors of $A$ is proportional to $(a+1)(a+5)$. At $a=-1$ and $a=-5$ the rank is three, while it is four everywhere else. The cohomology is therefore
\begin{equation}
\dim H^1\big|_{\ell=6,\,s=-6}=
\begin{cases}
2,&a=-1\ \text{or}\ a=-5,\\
0,&a\notin\{-1,-5\}.
\end{cases}
\label{eq:n6cohomology}
\end{equation}
Indeed, $\dim\ker D_1=9-\operatorname{rank}A$ and $\dim\operatorname{im}D_0=1+\operatorname{rank}A$. Thus the cohomology is enhanced at the two points $(\gamma_1,\gamma_2)=(-1,-5)$ and $(-5,-1)$, and vanishes at generic points of the line. At this level the other two projector sectors are contractible by (\ref{eq:Lpsi}), so (\ref{eq:n6cohomology}) gives the full ghost number one cohomology at level six on $s=-6$.

\bibliographystyle{utphys}
\bibliography{residual_kbc}

\end{document}